# BIN PACKING PROBLEM: TWO APPROXIMATION ALGORITHMS

Abdolahad Noori Zehmakan[1]

[1]Department of Computer Science, Sharif University of Technology, Tehran, Iran
`a_noori@mehr.sharif.ir`

*ABSTRACT*

*The Bin Packing Problem is one of the most important optimization problems. In recent years, due to its NP-hard nature, several approximation algorithms have been presented. It is proved that the best algorithm for the Bin Packing Problem has the approximation ratio 3/2 and the time order O(n), unless P=NP. In this paper, first, a $\frac{3}{2}$-approximation algorithm is presented, then a modification to FFD algorithm is proposed to decrease time order to linear. Finally, these suggested approximation algorithms are compared with some other approximation algorithms. The experimental results show the suggested algorithms perform efficiently.*
*In summary, the main goal of the research is presenting methods which not only enjoy the best theoretical criteria, but also perform considerably efficient in practice.*

## 1. INTRODUCTION

The Bin Packing Problem has several applications, including filling containers, loading trucks with weight capacity constraints, creating file backups in removable media and technology mapping in Field-programmable gate array semiconductor chip design. Unfortunately, this problem is NP-hard therefore many approximation algorithms [1, 2, 3, 4, 5] have been suggested.

In computer science and operational research, approximation algorithms are used to find approximate solutions to optimization problems. Approximation algorithms are often associated with *NP-hard* problems. They are also increasingly used for problems where exact polynomial-time algorithms are known but too expensive due to the input size. The quality and ability of an approximation algorithm depend on its approximation ratio and time order. For some approximation algorithms, it is possible to prove certain properties about the approximation of the optimal result. A ρ-approximation algorithm A is defined to be an algorithm for which it been proven that the value of the approximate solution A(x) to an instance x will not be more (or less, depending on the situation) than a factor ρ times the value, OPT, of an optimum solution.

In the classical one-dimensional Bin Packing Problem, a list of items $I = \{a_1, \ldots, a_n\}$, each with a size $s(a_i) \in (0,1]$, is given and we are asked to pack them into minimum number of unit-capacity bins.

Many variations of this problem is proposed, such as 2D and 3D bin packing [6, 7, 8, 9, 10], with item fragmentation [11], fragile objects [12, 13], extendable bins [14] packing by cost [3] and variable size bin packing [15]. In this paper, the original and off-line version of the problem is considered, due to its applications and importance.

Simchi-Levi in [16] proved that the FF (First-Fit) and BF (Best-Fit) algorithms, two of the foremost approximation algorithms for the Bin Packing Problem, have an absolute worst-case ratio of 7/4. He also proved that the *FFD* and BFD algorithms have an absolute worst-case ratio of 3/2. Zhang and Cai in [17] provided a linear time constant space off-line approximation algorithm with absolute approximation ratio of 3/2. Their algorithm depends on two kind of active and extra bins and follows a simple but exact procedure. In 2003, Rudolf and Florian in [18] presented an approximation algorithm for the BPP with a linear running time and an absolute approximation factor of 3/2. As mentioned, it is proven that the best algorithm for the Bin Packing Problem has the approximation ratio of 3/2 and the time order of $O(n)$, unless $P = NP$[16].

In [20] Martel defined the asymptotic approximation ratio instead of the approximation ratio and proved his proposed algorithm has a 4/3 asymptotic approximation ratio. Furthermore, in [20] the method of Martel was expanded and a 5/4 asymptotic approximation algorithm was suggested.

In this paper two new approximation algorithms are presented. The first algorithm works based on a kind of sorting and after classification items into 4 ranges tries to choose the best matching between them. The second algorithm is a time improved version of *FFD*. In this algorithm, we try to decrease *FFD* time order while maintaining the instructive qualities of *FFD* and its performance.

Finally, the two suggested algorithms are compared with two approximation algorithms [17, 18], and *FFD*. Experimental results show the two suggested algorithms perform much better than the others.

The reminder of this paper is organized as follows. In section 2, two suggested algorithms are presented. Furthermore, it is proved that the approximation factor of the first algorithm is 3/2. Then in sections 3 the experimental results and computational analysis are discussed. Finally, in section 4 conclusions of the results are drawn and some methods for enhancing previous algorithms are suggested.

## 2. THE PROPOSED ALGORITHMS

In this section, two proposed algorithm *A1* and *A2* are discussed. Algorithm *A1* utilizes ranging technique and classifies inputs into 4 ranges. It will be proved that this algorithm's approximation ratio is 3/2. Furthermore, a new linear version of *FFD* algorithm is presented.

### 2.1. The Proposed Algorithm A1

The algorithm tries to create output bins which are at least 2/3 full. It is proved that in this condition the approximation ratio of the algorithm is 3/2.

As mentioned, in this algorithm inputs are classified into 4 ranges $(0-\frac{1}{3})$, $(\frac{1}{3}-\frac{1.5}{3})$, $(\frac{1.5}{3}-\frac{2}{3})$ and $(\frac{2}{3}-1)$ called $S$, $M_1$, $M_2$ and $L$, respectively.

In first step, $L$ items are put in separate output bins, then $M_1$ and $M_2$ are sorted. We try to match any item in $M_2$ with the biggest possible item in $M_1$. Obviously, after that this step, some items will be remained in $M_1$ and $M_2$. We match $M_1$ items with each other and add $\frac{|M_1|}{2}$ to $Bin - counter$ (The number of used bins). In next step, we try to match $M_2$ items with $S$ items. Finally, $S$ items are matched with each other.

*Algorithm A1:*
*Read inputs & classify them into S, $M_1$, $M_2$, and L*
*Sort $M_1$ & $M_2$*
**For** *(any item $a$ in $M_2$)*
  **If** *($a$ can be matched with at least an item in $M_1$)*
    *Match $a$ with the biggest possible item in $M_1$;*
    *Eliminate them & Bin-counter ++;*
  **Else** *continue;*
*Bin-counter $+= \frac{|M_1|}{2}$;*
**For** *(any item $a$ in $M_2$)*
  **Do**
    *Choose an item $b$ in S;*
    *$a = a + b$ & eliminate $b$ & $c = b$;*
  **While** *($a \leq 1$)*
  *Eliminate $a$ & put $c$ in S & Bin-counter ++;*
**While** *(S is not empty)*
  *Choose an item $a$ in S;*
  **While** *( $a \leq 1$)*
    *Choose an item $b$ in S & $a = a + b$ & $c = b$;*
  *Eliminate $a$ & Bin-counter++ & put $c$ in S*
**End**

**Definition1**: $P$ is the number of bins in OPT solution and $P^*$ is the number of bins in the proposed algorithm.

**Lemma1**: If at least $\frac{2}{3}$ size of each output bin is full, the approximation ratio is at least $\frac{3}{2}$.

**Proof**: consider the worst condition that all output bins are completely full in OPT solution. Suppose that $W$ is the sum of input items. In this condition:
$$P \geq W \ \& \ P^* \leq \frac{W}{\frac{2}{3}} \Rightarrow P^*/P \leq \frac{3}{2} \quad \blacksquare$$

**Theorem1**: The proposed algorithm *A1* is a $\frac{3}{2}$-approximation algorithm.

**Proof**: Based on the algorithm in first step, all $L$ items are put in separated bins and obviously at least 2/3 size of these output bins are full. After that, some $M_2$ items are matched with some $M_1$ items. Definitely, in this step at least 2/3 size of output bins are also full since a $M_1$ item is at least 1/3 and a $M_2$ item is at least 1.5/3. Consequently, their sum is at least 2/3.

In next step, $M_1$ items are matched with each other 2 by 2 and put in separated bins. At least 2/3 size of these bins are full since an $M_1$ item is at least 1/3. After that the rest of $M_2$ items with $S$ items are matched. Now there are two cases:

Case1: $\frac{W_{M_2} + W_S}{|M_2|} > \frac{2}{3}$

Case2: $\frac{W_{M_2} + W_S}{|M_2|} \leq \frac{2}{3}$

$W_{M_2}$ : *The sum of all $M_2$ items which remain in this step.*
$W_S$ : *The sum of all S items.*
$|M2|$ : *The number of all $M_2$ items which are remain in this step.*

We claim all output bins are more than $\frac{2}{3}$ fill in this step. According on the algorithm, at first we match some $M_2$ items with some $S$ items. Obviously the output bins in this step are more than $\frac{2}{3}$ full because a $S$ item is no more than $\frac{1}{3}$ and we close a bin when it does not have enough space for a S item. After that, two configurations are possible:

C1: If there are just some $M_2$ items left we put all of them into separate bins therefore the number of output bins is $|M2|$. Consequently the average of the output bins equals $\frac{W_{M_2}+W_S}{|M_2|}$ that is more than $\frac{2}{3}$ based on case1 assumption.

C2: If there are only some $S$ items, the output bins in this step are also more than $\frac{2}{3}$ full because a $S$ item is at most $\frac{1}{3}$.

In case2, the bins that have some S items like case1 are at least $\frac{2}{3}$ full. Therefore we only consider the bins which have only one $M_2$ item. We claim that in the OPT solution these $M_2$ items are also associated separate bins because:

On one hand, they cannot be matched with the $L$ items and with the $M_2$ items because a bin does not have enough space for an $L$ item and an $M_2$ item or for two $M_2$ items. On the other hand, if a $M_2$ item (primary $M_2$ item) is matched with a $M_1$ item in the OPT solution, in the suggested algorithm it will be matched with a $M_1$ item or its complement (meaning the $M_1$ item matched with it in the OPT solution) is matched with another $M_2$ item (second $M_2$ item). The second $M_2$ item is bigger than the primary $M_2$ item since the $M_2$ items are sorted. Therefore, the primary $M_2$ item can be put in every bin that the second $M_2$ has been put (in this condition the algorithm has been performed better than OPT solution until now).

Based on the mentioned reasons and discussions, for any output bin in the proposed algorithm which is less than $\frac{2}{3}$ full, there is a bin in the OPT solution that its used capacity is equal or less than it. Furthermore, all other bins are more than $\frac{2}{3}$ full. In conclusion, based on the *lemma1* the approximation ratio of the suggested algorithm is $\frac{3}{2}$. ∎

## 2.2. The Proposed Algorithm A2

As mentioned, the second proposed algorithm is based on the *Firs-Fit Decreasing* algorithm. In *FFD*, the items are packed in order of non-decreasing size, and next item is always packed into the first bin in which it fits; that is, we first open bin1 and we only start bin k+1 when the current item does not fit into any of the bins $1, \dots, k$.

In the algorithm $A2$, we consider 10 classes of bins and 10 ranges of items and in any step we check at most one bin in each class. The order of choosing items and checking the bins classes are considered completely intelligently. A pseudocode of the algorithm $A2$ is shown.

> *Algorithm A2:*
> **Consider** *10 sets of bins $B_i \, \forall \, 0 \leq i \leq 9$. A bin is in the set $B_i$ if the bin's free space is between ($i*0.1$) and (($i+1$) $0.1$). (At first all sets are empty;*
> **Consider** *10 ranges $R_i = (i * 0.1, (i+1) * 0.1) \, \forall \, 0 \leq i \leq 9$.*
> **Read** *items from input.*
> **Put** *the items into corresponding ranges*
> **For** *($i = 9; \, i > -1; \, i--$)*
>     **While** *($R_i$ is not empty)*
>         **Choose** *an item $a$ in $R_i$ randomly;*
>         **For**($j = 0; \, j < 10; \, j++$)
>             **Choose** *a random bin in $B_j$;*
>             **If** *(the bin has enough space for $a$)*
>                 **Put** *$a$ in the bin & change that to the appropriate $B_i$;*
>                 **Eliminate** *$a$ from $R_i$;*
>                 **Break**;
>         **Put** *the item $a$ into an empty bin and put the bin into corresponding $B_i$;*
>         **Eliminate** *$a$ from $R_i$*

Obviously, the running time of the algorithm *A2* is $O(n)$ (n is the number of input items) since for making decision about each item the algorithm at most spend 10 time-unit for checking 10 classes of bins.

We also can make the algorithm more efficient and consider the *Scale Parameter r* that shows the number of ranges and bins classes in the algorithm. This parameter can be chosen based on the number of inputs. For instance, if the number of inputs is $10^{10}$ is reasonable choose $r = 10^3$ instead of $r = 10$.

## 3. COMPUTATIONAL RESULTS

In this section, at first the computational results of two suggested algorithms and three other algorithms are presented, and it is shown that the proposed algorithms perform considerably much more efficient. Furthermore, we compare the algorithm *A1* with the Algorithm *A2* from an application point of view and their utilization in variant fields and stipulations.

In this section, the two proposed algorithms are compared with two other approximation algorithms [18, 19] which are the only algorithms have the best possible approximation ratio. This comparison has been drawn based on all standard instances for BPP from **OR-LIBRARY** [21]. We define *Ratio* as the proportion of the proposed algorithm solution to the OPT solution. Obviously, *ratio* has a direct relationship with algorithm's approximation ratio. Consequently, *ratio* is utilized as a factor for measuring approximation algorithms' performances.

As mentioned, the standard instances in OR-LIBRARY are used for simulations. Each set of instances contains 20 instances for the Bin Packing Problem. The two proposed algorithm have been compared with the *Guochuan's* algorithm [17], and the *Berghammer*'s algorithm [18] based on the 8 set of instances. The results of these comparisons for *bp1, bp2, bp3, bp4, bp5, bp6, bp7* and *bp8* are shown in Fig1, Fig2, Fig3, Fig4, Fig5, Fig6, Fig7, Fig8, respectively.

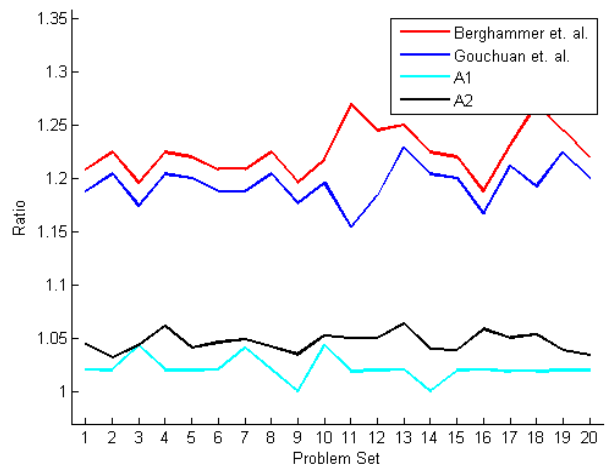

Figure 1. The ratios of the algorithms for the set problems of instance bp1

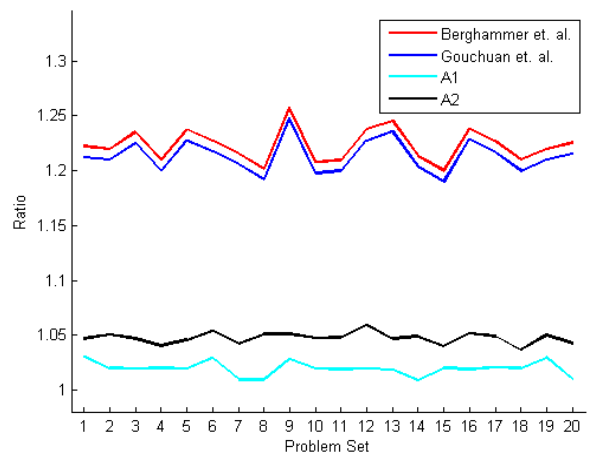

Figure 2. The ratios of the algorithms for the set problems of instance bp2

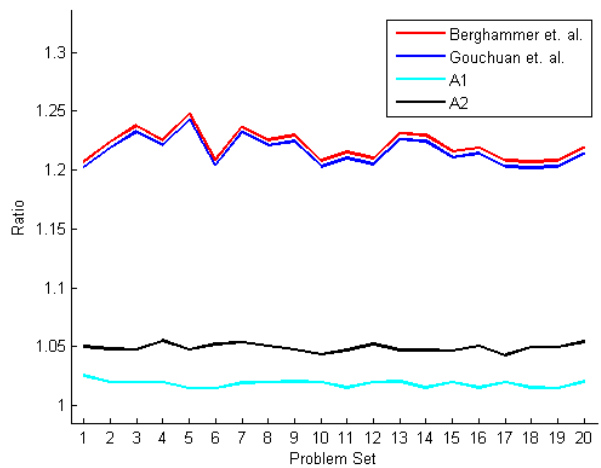

Figure 3. The ratios of the algorithms for the set problems of instance bp3

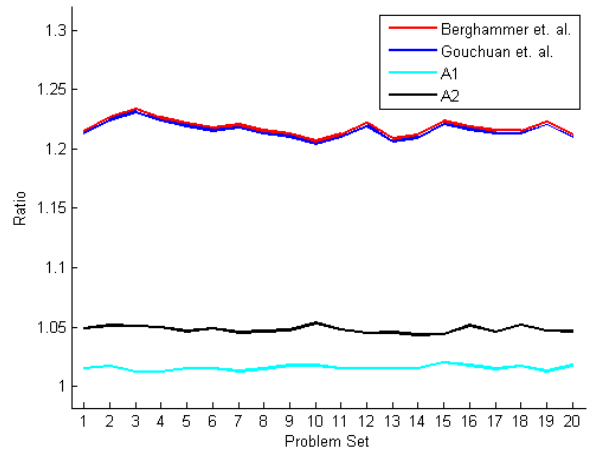

Figure 4. The ratios of the algorithms for the set problems of instance bp4

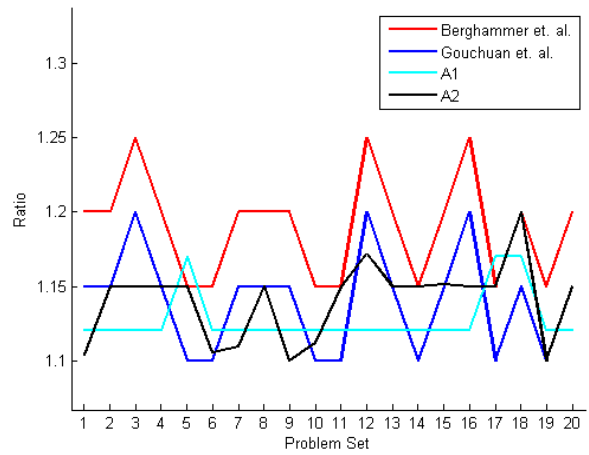

Figure 5. The ratios of the algorithms for the set problems of instance bp5

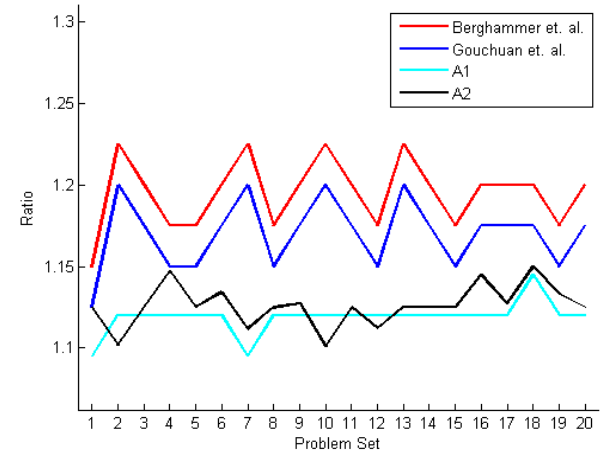

Figure 6. The ratios of the algorithms for the set problems of instance bp6

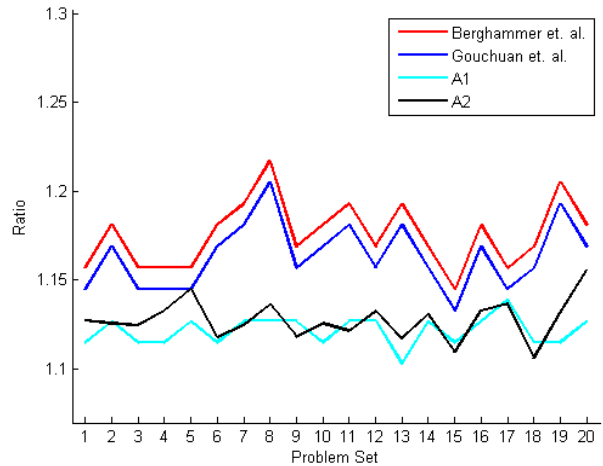

Figure 7. The ratios of the algorithms for the set problems of instance bp7

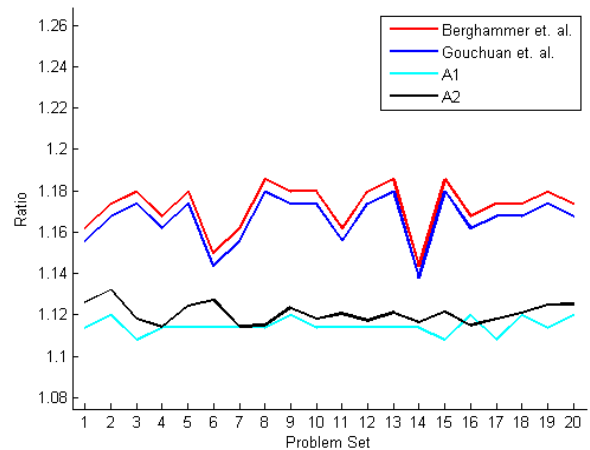

Figure 8. The ratios of the algorithms for the set problems of instance bp8

The diagrams show the two suggested algorithms perform much better than two other algorithms. As mentioned, the two other algorithms are only approximation algorithms with the best possible approximation factor. Furthermore, the algorithm *A1* performance is more acceptable than the algorithm *A2*. Another interesting point in the experimental results is the similarity between performances of *Guochuan's* algorithm, and the *Berghammer*'s algorithm. The results are measured for 20 instances in any class, but for simplification of understanding the points corresponding to an algorithm are joined by a line.

In Fig9, the average of the simulations results is shown for four mentioned algorithms for the all sets of instances. This diagram shows that the proposed algorithm *A1* in all instances performs more efficiently. After that, the suggested algorithm *A2* has much better performance. Therefore, two suggested algorithms are completely superior to two other ones, in practice.

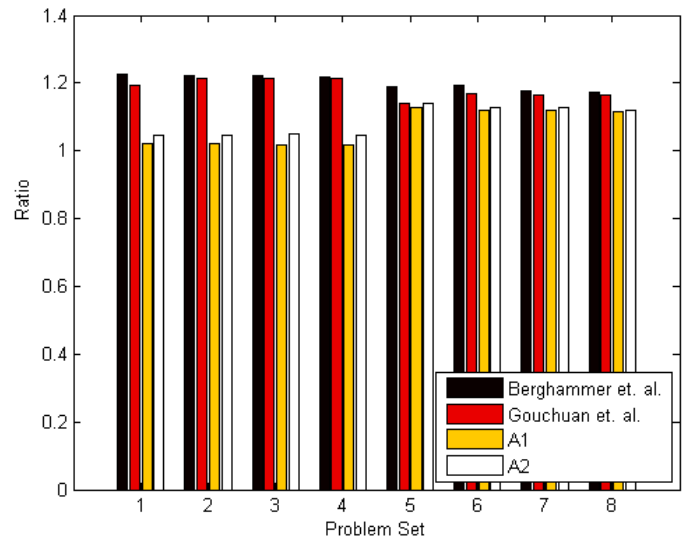

Figure 9. The average of ratios for the 4 algorithms based on the all instances

In Fig 10, the experimental results of the two suggested algorithms and *FFD* algorithm are shown based on the all sets of instances.

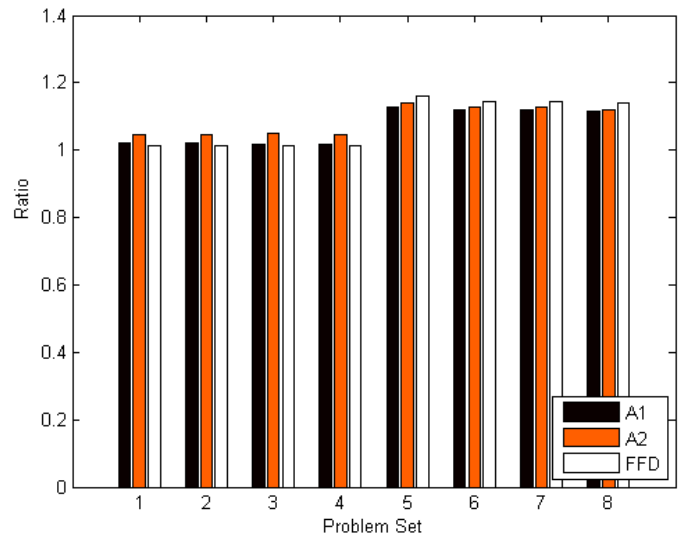

Figure 10. The average of ratios for two suggested algorithms and FFD based on the all instances

The results show that the two suggested algorithms perform much better than *FFD* algorithm in bp5, bp6, bp7, and bp8, but the *FFD* algorithm performances are more acceptable in bp1, bp2, bp3 and bp4. It seems their performances are very similar in average. We claim that the two suggested algorithms are more effective and efficient than *FFD*. The algorithm *A1* and *FFD* time order are similar, but *FFD* is an on-line space algorithm (it means that it save all bins during the algorithm) while the algorithm *A1* use much less space. Furthermore, the algorithm *A2* is also superior to *FFD* because it is a linear time algorithm while the running time of *FFD* is $O(nlogn)$ even in worst-case $O(n^2)$.

We drew the conclusion that the algorithms *A1* and *A2* not only enjoy the best possible theoretical criteria, but also execute better than other ones in practice, but a natural question which comes up is that "Which algorithm should be used in practice, *A1* or *A2*?". The answer is that it depends. In the following paragraphs we try to clarify this point.

Firstly, obviously if the important factor is accuracy, *Algorithm A1* is the better one, but if the significant criterion is speed, *Algorithm A2* will be the choice inasmuch as *Algorithm A1* shows better performance based on the aforementioned outputs; on the other hand, *Algorithm A2* is a linear time algorithm. Another point which can be taken into consideration is that *Algorithm A1* is a constant-space one while the second one is not. Therefore, if space order is a noteworthy factor, we should exploit *Algorithm A1*.

Needless to say, if the input items are almost sorted, the algorithm *A1* performs a lot better, but if the number of input items is significantly high or they are distributed homogenously, the algorithm *A2* will be the option In that *Algorithm A1* needs to sort the items, and the algorithm *A2* is much more flexible and is able to use *Scale Factor*. The aforementioned computational results confirm this claim because the number of items in the instances increases from *bp1* to *bp8*.

If the number of *S* (small) items is considerable, Algorithm *A1* performs more efficiently. On the other hand, if the number of *L* (large) items is high, the second one is the right choice. Moreover, the state that nearly all items are relevant to the ranges *M1* and *M2* (are medium) forces the user to utilize the algorithm *A2*.

For instance, in packing trucks and ships when the goods are small, we use the first one, but in the state that they are large enough by considering the capacity unit in the ship or truck, the choice is second one. Furthermore, in assigning tasks to machines in *machine scheduling problem* if the durations of different tasks are approximately equal with each other, the second algorithm executes better.

Consider the problem of placing computer files with specified sizes into memory blocks of fixed size. For example, recording all of a computer's music where the length of the pieces to be recorded are the weights and the bin capacity is the amount of time that can be sorted on an audio (say 80 minutes). If we want to save the information for a long time, it is better to use the first algorithm to amplify the accuracy, but if we want to rewrite the information several times, using the second one is a rational solution. If all items are similar in size, for instance all of them are songs, probably *Algorithm A1* works acceptably.

*Table 1* tries to summarize the aforementioned discussions regarding the application of the algorithms *A1* and *A2* in different situations.

Table 1. Choosing between algorithms 1 and 2 based on different factors and condition.

| Factor/Condition | Algorithnm1 | Algorithm2 |
|---|---|---|
| *Accuracy* | Yes | |
| *Speed* | | Yes |
| *Space* | Yes | |
| *Sorted Items* | Yes | |
| *High Number of Items* | | Yes |
| *Homogenous Distribution of Items* | | Yes |
| *Majority by S Items* | Yes | |
| *Majority by L Items* | | Yes |
| *Majority by M Items* | Yes | |

## 3. CONCLUSIONS

Two approximation algorithms *A1*, and *A2* were proposed in this paper. It was proved that the *A1* approximation ratio is $\frac{3}{2}$. After that we observed the results of experimental simulations and analyzed them. Based on the results, we can claim that the two proposed algorithms in this article are the best presented approximation algorithms for the Bin Packing Problem, in theory and in practice until now.

In future researches, the focus on *Scaling Factor r* can enhance the algorithm *A2* more and more.